\documentclass[aps,pra,preprint,groupedaddress,showpacs]{revtex4}
\begin{document}

\title{Quantum mechanics from two physical postulates}

\author{Mohammad Mehrafarin}
\affiliation{Physics Department, Amirkabir University of Technology, Tehran 15914,
Iran}
\email{mehrafar@aut.ac.ir}

\date{\today}

\begin{abstract}
For an arbitrary preparation, quantum mechanical descriptions refer to the complementary contexts set by incompatible measurements. We argue that an arbitrary preparation, therefore, should be described with respect to such a context by its degrees of disturbance (represented by real numbers) and their probability distribution (postulate 1). Measurement contexts thus provide reference frames for the preparation space of a physical system; a preparation being described by a point in this space with the aforementioned as its coordinates relative to a given measurement apparatus. However, all measurement contexts are equivalent with regard to the description of a given preparation; there is no preferred measurement (postulate 2). In the framework provided by the preparation space, we show that quantum mechanics emerges naturally from the above postulates in a new formulation which is manifestly canonical; provided the degrees of disturbance are identified with the quantum phases of the preparation with respect to (the basis furnished by) the measurement apparatus.
\end{abstract}


\maketitle

\section{Introduction: The postulates}
Among various estimates $\{p_i+dp_i\}$ for the probability distribution $\{p_i\}$; where $i=1,...,n$; that estimate is best which makes the quantity,
\begin{equation}
\sum_{i} \frac{dp_i^2}{4p_i}= \sum_{i} (d\sqrt{p_i})^2, 
\end{equation}
least (the $\chi^2$ criterion \cite {Jaynes}). This quantity, which has been referred to as the statistical distance \cite{Fisher, Bhattacharyya, Wootters}, therefore, provides a natural measure of distinguishability between two neighboring probability distributions $\{p_i\}$ and $\{p_i+dp_i\}$. It is, therefore, a measure of discrimination between the distributions of the various results (labeled by $i$) for two slightly different preparations under measurement \cite{Wootters}. The statistical distance, thus, furnishes a measure of the number of distinguishable preparations between two such preparations by the measurement apparatus, provided an arbitrary preparation can be described (with respect to the apparatus) by the probability distribution of its outcome $\{p_i\}$ {\it alone}. Such a description is, however, valid only in the classical limit according to the following argument.

For an arbitrary preparation, quantum mechanical descriptions refer to the complementary contexts set by incompatible measurements and it is with respect to such a context that preparations are described. (Mathematically, different bases (representations) for the Hilbert space are provided by incompatible measurements through the eigenstates of the measured observable; a change of basis corresponding to a change of measurement context (measurement apparatus); and state vectors (preparations) are described with respect to such a basis by complex components.) In view of the complementarity of contexts, the description of a preparation with respect to a measurement apparatus by $\{p_i\}$ alone is not sufficient to address the context, because the preparation may undergo two incompatible measurements with the same probability distribution of the outcome. For example, consider a measurement in which a single particle, prepared in a definite momentum eigenstate, is incident on a beam splitter that sends the particle along one of the two arms. The particle is then collected by means of two detectors, one facing each arm, whose counts constitute the measurement results. Consider, also, a different (incompatible) measurement in which the split beams are mixed via a second beam splitter before being collected by the detectors (the Mach-Zehnder setup). By adjustment of the path difference in the second measurement, the probability distribution of the detector counts could be made to coincide with that of the first measurement, namely, $\{\frac{1}{2},\frac{1}{2}\}$. We would then have a preparation undergoing two incompatible measurements with the same probability distribution of the outcome, and a description in terms of the latter alone, thus, fails to discriminate between the two (complementary) measurement contexts. However, in the classical limit, in view of the compatibility of all measurements, the difference (complementarity) of contexts disappears and the description of an arbitrary preparation, hence, becomes `non-contextual'; allowing it to be given in terms of the distribution of the outcome $\{p_i\}$ alone. 

The physical roots of incompatibility can be traced to the disturbance of the system by the act of measurement; incompatible measurements disturb a given preparation to different extents. The disturbance, which is the result of complex physical causations operating within the setup, if quantified, thus, provides a measure to discriminate between complementary contexts even when the distributions of the results are the same. In the above example, the degrees of disturbance of the preparation by the two measurement apparatus, being different in each case, therefore, could be used to discriminate between the two measurement contexts. Hence, in the description of an arbitrary preparation with respect to a given measurement context (apparatus) we propose that, in addition to $\{p_i\}$, the degree of disturbance be taken also into account in order to have a contextual description that addresses the situation fully. (In the classical limit, all measurements are compatible because the degrees of disturbance of the preparation by all measurement apparatus are the same (infinite) and, hence, irrelevant to the description of the preparation.) 

To fulfill the above desideratum we obviously need some kind of association between the degrees of disturbance and real numbers. Now, in every repetition of a measurement, the randomness of the result (an eigenvalue of the measured observable) is attributed to the uncontrollable disturbance of the preparation by the measurement apparatus. Thus, let $q_i$ be a real-valued measure of the (one-to-one) relation between the result $i$ and the random disturbance. In other words, $q_i$ shall represent the degree of disturbance of the preparation in any repetition that produces the result $i$. Then, the random variable with the set of possible values $\{q_i\}$ is a measure of the degree of disturbance of the preparation by the act of measurement. Clearly, the probability distribution of this random variable coincides with that of the results, $\{p_i\}$, and that the two sets of values $\{q_i\}$ and $\{p_i\}$ are independent. Of course, the absolute value of $q_i$ is arbitrary \cite{note}. However, the relative value $q_{ij}=q_i-q_j$ is physically significant, because it measures the difference in the degrees of disturbance of the preparation corresponding to the different results $i$ and $j$. As a consequence, $\{q_i\}$ can be specified only up to a common constant; i.e., two sets of values $\{q_i\}$ and $\{q_i+const.\}$ with the same probability distribution $\{p_i\}$, represent the same disturbance. 

The collection of values $(p_i,q_i)$, therefore, specifies an arbitrary preparation with respect to a given measurement context (apparatus). The totality of all preparation states of a given physical system is called the preparation space of the system and an arbitrary preparation can, hence, be represented by a point in this space with coordinates $(p_i,q_i)$ relative to a given measurement context. Measurement apparatus, thus, provide reference frames or coordinate systems for the preparation space relative to which a point (preparation) is described by its degrees of disturbance and their probability distribution (which is also the probability distribution of the results) as its coordinates. However, every reference frame is equivalent to every other frame with regard to the description. We, thus, raise the following basic postulates that form the framework of the preparation space.

Postulate 1: {\it An arbitrary preparation is described with respect to a    measurement context by its degrees of disturbance and their probability distribution.}

Postulate 2: {\it All measurement contexts are equivalent with regard to the description of a given preparation; there is no preferred measurement.}

We shall see how quantum mechanics emerges naturally in the preparation space from the above two postulates in a new formulation that is manifestly canonical. 

\section{Preparation space: The line element and the transformation law}
The disturbance of a system by a given measurement apparatus depends on its state of preparation. In distinguishing between two neighboring preparations in the preparation space by a measurement apparatus, in addition to discriminating between the corresponding probability distributions (the contribution of the classical statistical distance (1)), their degrees of disturbance have to be discriminated too. Bearing in mind that $\{q_i\}$ can be specified only up to a common constant, among various estimates $\{q_i+d q_i\}$ for $\{q_i\}$ with a given probability distribution $\{p_i\}$, that estimate is best which makes the variance,
\begin{equation}
\sum_{i} p_i dq_i^2-(\sum_{i} p_i dq_i)^2, 
\end{equation}
least. (This is essentially the least-square criterion.) Hence, (2) provides a natural measure of discrimination between two sets of values $\{q_i\}$ and $\{q_i+d q_i\}$ with the same probability distribution $\{p_i\}$; i.e., it provides a measure of discrimination (with respect to a given apparatus) between the degrees of disturbance of two neighboring preparations. The variance (2), therefore, represents the distance between two neighboring preparations at fixed $\{p_i\}$, in the same way that (1) represents the distance at fixed $\{q_i\}$. With respect to a given measurement apparatus, the number of distinguishable preparations between two arbitrary neighboring preparations $(p_i,q_i)$ and $(p_i+dp_i, q_i+dq_i)$ can, therefore, be represented by the (Riemannian) line element,
\begin{equation}
ds^2=\sum_{i} \frac{dp_i^2}{4p_i}+ \sum_{i} p_i dq_i^2-(\sum_{i} p_i dq_i)^2,
\end{equation}
of the preparation space. Line element (3) reduces to the classical statistical distance (1) in the limit where all measurements are compatible so that the contribution of the measure of discrimination between the degrees of disturbance (the variance term), being the same with respect to all reference frames (measurement apparatus), can be discarded.

Now, let $(p_i^\prime,q_i^\prime)$ and $(p_i^\prime+dp_i^\prime, q_i^\prime+dq_i^\prime)$ be the coordinates of the same two neighboring preparations with respect to a different  reference frame (measurement apparatus). Then with respect to that frame, the number of distinguishable intermediate preparations is given by, 
\begin{equation}
ds^{\prime 2}=\sum_{i} \frac{dp_i^{\prime 2}}{4p_i^\prime}+ \sum_{i} p_i^\prime dq_i^{\prime 2}-(\sum_{i} p_i^\prime dq_i^\prime)^2.
\end{equation}
However, because there is no preferred frame (postulate 2), the measure of distinguishability has to be the same with respect to all measurement apparatus; i.e., $ds^{\prime 2}=ds^2$. (Otherwise, some measurements would be more discriminating than others which provides a basis for preference, violating postulate 2 \cite{note2}.)

The invariance of the line element restricts the form of the allowed coordinate transformations $(p_i,q_i) \rightarrow (p_i^\prime,q_i^\prime)$ in the preparation space. Such transformations will determine how a given preparation is to be described with respect to different measurement contexts. Of course, $\sum_{i} p_i$ must remain invariant under the transformations for the normalization condition to be preserved. Introducing the `Cartesian-like' variables,
$$x_i= \sqrt {p_i}\ \cos q_i, \ \ \ \ y_i= \sqrt {p_i}\ \sin q_i,$$
we have,
\begin{eqnarray}
ds^2=\sum_{i} (dx_i^2+dy_i^2)-[\sum_{i} (x_i dy_i-y_i dx_i)]^2,\\
\sum_{i} p_i= \sum_{i}(x_i^2+y_i^2). \ \ \ \ \ \ \ \ \ \ \ \ \ \ \ \ \ 
\end{eqnarray}
We seek transformations $(x_i,y_i) \rightarrow (x_i^\prime,y_i^\prime)$ that leave (5) and (6) simultaneously invariant. The most general (linear) transformations are,
$$x_i^\prime= \sum_{j} (a_{ij} x_j+b_{ij} y_j),$$
$$y_i^\prime= \sum_{j} (c_{ij} x_j+d_{ij} y_j),$$
where the $4n^2$ coefficients are all constant (independent of $x_i$ and $y_i$). The invariance of (6) imposes the constraints,
\begin{equation}
\sum_{i} (a_{ij} a_{ik}+c_{ij} c_{ik})=\sum_{i} (b_{ij} b_{ik}+d_{ij} d_{ik})= \delta_{jk}, \ \ \ 
\sum_{i} (a_{ij} b_{ik}+c_{ij} d_{ik})=0.
\end{equation}
These are $(2n^2+n)$ independent equations leaving, as they should, only $(2n^2-n)$ of the transformation parameters free. Moreover, bearing in mind that the first summation in (5) has now become invariant too, the invariance of the line element, therefore, requires that the second summation be also independently invariant under the transformation. This, then, introduces $(2n^2-n)$  more constraints, namely, 
\begin{equation}
\sum_{i} (a_{ij} c_{ik}-c_{ij} a_{ik})=\sum_{i} (b_{ij} d_{ik}-d_{ij} b_{ik})= 0,\ \ \ 
\sum_{i} (a_{ij} d_{ik}-c_{ij} b_{ik})=\delta_{jk}.
\end{equation}
It is easy to show that the two sets of constraints (7) and (8) are consistent if and only if, 
$$c_{ij}=-b_{ij}, \ \ \ \ d_{ij}=a_{ij}.$$
Hence, the most general transformation that leaves the line element and the normalization condition simultaneously invariant is of the form,
\begin{eqnarray}
x_i^\prime= \sum_{j} (a_{ij} x_j+b_{ij} y_j),\ \nonumber \\
y_i^\prime= \sum_{j} (-b_{ij} x_j+a_{ij} y_j), 
\end{eqnarray}
where the $2n^2$ coefficients satisfy,
\begin{equation}
\sum_{i} (a_{ij} a_{ik}+b_{ij} b_{ik})= \delta_{jk}, \ \ \ 
\sum_{i} (a_{ij} b_{ik}-b_{ij} a_{ik})=0.
\end{equation}
These are $n^2$ constraints leaving only $n^2$ transformation parameters independent. The inverse transformation is then given by,
$$x_i= \sum_{j} (a_{ji} x_j^\prime-b_{ji} y_j^\prime),$$ 
$$y_i= \sum_{j} (-b_{ji} x_j^\prime+a_{ji} y_j^\prime),$$
and the same constraints can be written also as,
\begin{equation}
\sum_{i} (a_{ji} a_{ki}+b_{ji} b_{ki})= \delta_{jk}, \ \ \ 
\sum_{i} (a_{ji} b_{ki}-b_{ji} a_{ki})=0.
\end{equation}
Returning to the `polar-like' variables $(p_i,q_i)$, let us write,
$$a_{ij}= \sqrt {w_{ij}}\ \cos \beta_{ij}, \ \ \ \ b_{ij}= \sqrt {w_{ij}}\ \sin \beta_{ij}.$$
Using (10) and (11), the $n^2$ conditions on the new transformation parameters $w_{ij}$ and $\beta_{ij}$, thus, translate into,
\begin{eqnarray}
\sum_{i} \sqrt {w_{ij}w_{ik}}\ \matrix {\cos\cr \sin\cr}(\beta_{ik}-\beta_{ij})&=& \sum_{i} \sqrt {w_{ji}w_{ki}}\ \matrix {\cos\cr \sin\cr} (\beta_{ki}-\beta_{ji})=0, \ \ (j \neq k) \nonumber \\
\sum_{i} &w_{ij}&= \sum_{i} w_{ji}=1. 
\end{eqnarray}
As for the coordinate transformation, (9) becomes,
\begin{equation}
\sqrt{p_i^\prime}\ \matrix {\cos\cr \sin\cr} q_i^\prime= \sum_{j} \sqrt{w_{ij}p_j}\ \matrix {\cos\cr \sin\cr} (q_j-\beta_{ij}),
\end{equation}
i.e.,
\begin{eqnarray}
p_i^\prime= \sum_{jk} \sqrt{w_{ij}p_j}\ \sqrt{w_{ik}p_k}\ \cos (q_{jk}-\beta_{ij}+\beta_{ik}), \nonumber \\
\tan q_i^\prime= \frac {\sum_{j} \sqrt{w_{ij}p_j}\ \sin (q_j-\beta_{ij})}{\sum_{j} \sqrt{w_{ij}p_j}\ \cos (q_j-\beta_{ij})}.\ \ \ \ \ \
\end{eqnarray}
Writing the first equation as,
\begin{equation}
p_i^\prime= \sum_{j} w_{ij}p_j + \sum_{j \neq k} \sqrt{w_{ij}p_j}\ \sqrt{w_{ik}p_k}\ \cos (q_{jk}-\beta_{ij}+\beta_{ik}),
\end{equation}
and using (12), it can be readily checked that $\sum_{i} p_i^\prime= \sum_{i} p_i$. Through is random behavior, the disturbance is responsible for the mysterious `interference' effects represented by the second summation in (15). In the classical limit of infinite disturbance $q_i \rightarrow \infty$, this term vanishes due to the infinitely rapid oscillations of the cosine, yielding the standard probability rule for mutually exclusive results in terms of the conditional probabilities $w_{ij}$. This reconciles with the fact that in this limit, because all measurements are compatible, the disturbance becomes irrelevant and the classical non-contextual description is restored.
 
Coordinate transformation (14) is the required transformation law in the preparation space, which relates, in terms of $n^2$ independent parameters, the descriptions of a given preparation with respect to different measurement contexts. 

\section{Correspondence with quantum mechanics}
In the standard formulation of quantum mechanics, different bases (representations) for the Hilbert space are provided by incompatible measurements through the eigenstates $\{|i>\}$ of the measured observable. A state vector $|\psi>$, which represents a preparation, is described with respect to such a basis by complex components $\psi_i= \sqrt{p_i}\ e^{i \phi_i}$; $\phi_i$ being the quantum phase of the preparation corresponding to the result $i$ of the measurement. A change of basis (representation) $\{|i>\} \rightarrow \{|i^\prime>\}$, thus, corresponds to a change of measurement context and is accompanied by a complex transformation matrix $u_{ij}=<i|j^\prime>$, which; being unitary; also involves $n^2$ independent (real) parameters. Indeed, writing
$$u_{ji}= \sqrt {w_{ij}}\ e^{i \beta_{ij}},$$
the $n^2$ unitary conditions, namely,
$$\sum_{i} u_{ij}^* u_{ik}= \sum_{i} u_{ji} u_{ki}^*= \delta_{jk},$$
simply translate into equations (12). That is, any unitary matrix can be written in the above form where $w_{ij}$ and $\beta_{ij}$ satisfy conditions (12). Hence, the transformation parameters $w_{ij}$ and $\beta_{ij}$ involved in the change of measurement apparatus correspond to the unitary transformation matrix of the standard formulation. Now, under a change of basis, the components $\psi_i$ of the state vector transform as
$$\psi_i^\prime= \sum_{j} u_{ji}^* \psi_j.$$
But this is just the transformation law (13), provided we identify $q_i$ with $\phi_i$. In other words, the quantum phases of a given preparation with respect to (the basis provided by) a measurement apparatus are just the degrees of disturbance of that preparation by the apparatus. The irrelevance of an overall phase factor then complies with the fact that the absolute value of the degree of disturbance is arbitrary.
 
Furthermore, the natural measure of distinguishability between two state vectors (preparations) in the Hilbert space is provided by the angle between the corresponding rays. For two neighboring preparations, the angle is,
\begin{eqnarray}
cos^{-1}  |<\psi|\psi+d \psi>|= cos^{-1} |\sum_{i} \sqrt{p_i(p_i+dp_i)}\ e^{i d\phi_i}| \nonumber \ \ \\
= \sum_{i} \frac{dp_i^2}{4p_i}+ \sum_{i} p_i d\phi_i^2-(\sum_{i} p_i d\phi_i)^2+higher\ order\ terms. \nonumber
\end{eqnarray}
This corroborates expression (3) for the line element of the preparation space. Because unitary transformations are angle preserving, it is not surprising why the unitary group of quantum mechanics has emerged as the symmetry group of the line element, namely, the group of transformations defined by (14). 

Let us recapitulate the foregoing. For an arbitrary preparation, quantum mechanical descriptions refer to the complementary contexts set by incompatible measurements. With respect to such a context, we argued that the preparation should be described by its degrees of disturbance (represented by real numbers) and their probability distribution. The former, notably, enters the description because it is contextual, i.e., it depends (in view of incompatibility) on the measurement apparatus. The latter, due to the one-to-one correspondence between the disturbance at each repetition and the measurement result, simply coincides with the probability distribution of the results. The above desideratum is fulfilled via postulate 1. Measurement apparatus, thus, provide reference frames for the preparation space of a physical system, relative to which a point (preparation) is described by its degrees of disturbance and their probability distribution as its coordinates. Thence, in distinguishing between two neighboring preparations by a measurement apparatus, in addition to discriminating between the corresponding probability distributions (the contribution of the classical statistical distance (1)), their degrees of disturbance ought to be discriminated too. This leads, naturally, to the line element (3) of the preparation space for the measure of distinguishability between two arbitrary neighboring preparations. Nevertheless, in spite of the contextuality of quantum mechanical descriptions, there is no preferred measurement for the description of a given preparation. This desideratum is fulfilled by postulate 2, which implies that the line element is invariant with respect to all measurement contexts. Whence, the unitary transformation group of quantum mechanics emerges as the group of coordinate transformations (14) (the symmetry group of the line element); which relates the descriptions of the preparation with respect to different measurement apparatus; provided we identify the degrees of disturbance of the preparation (by the apparatus) with its quantum phases (with respect to the apparatus). The complex (Hilbert) space structure of quantum mechanics, therefore, emerges from two physical postulates; most notably postulate 1, through which the meaning of phase and its physical role in quantum mechanics are unveiled. In the following section, we show that the evolution law of an isolated preparation (the Shr\"{o}dinger equation) naturally follows, too, from the same principles.

\section{Dynamics in the preparation space: The canonical equations of motion}
In the preparation space of an isolated system, consider an arbitrary preparation specified by the coordinates $(p_i,\phi_i)$ with respect to a given measurement apparatus. Because there is no preferred frame, the equations governing the time development of the preparation have to be covariant with respect to the transformation law (14). Now from (14),
$$\frac {\partial p_i^\prime}{\partial p_j}= \frac{C_{ij}}{p_j},\ \ \ \ \frac {\partial p_i^\prime}{\partial \phi_j}= -2 S_{ij},$$
$$\frac {\partial \phi_i^\prime}{\partial p_j}= \frac{1}{2p_j}\ \frac{S_{ij}}{\sum_{k} C_{ik}},\ \ \ \ \frac {\partial \phi_i^\prime}{\partial \phi_j}= \frac{C_{ij}}{\sum_{k} C_{ik}},$$
where,
$$\matrix {C_{ij}\cr S_{ij}\cr}= \sum_{k} \sqrt{w_{ij}p_j}\ \sqrt{w_{ik}p_k}\ \matrix {\cos\cr \sin\cr} (\phi_{jk}-\beta_{ij}+\beta_{ik}).$$
It follows after some calculations, using conditions (12), that,
$$\sum_{j} (\frac {\partial p_i^\prime}{\partial p_j} \frac {\partial \phi_k^\prime}{\partial \phi_j}- \frac {\partial p_i^\prime}{\partial \phi_j} \frac {\partial \phi_k^\prime}{\partial p_j})= \delta_{ik}$$
$$\sum_{j} (\frac {\partial p_i^\prime}{\partial p_j} \frac {\partial p_k^\prime}{\partial \phi_j}-\frac {\partial p_i^\prime}{\partial \phi_j} \frac {\partial p_k^\prime}{\partial p_j})= \sum_{j} (\frac {\partial \phi_i^\prime}{\partial p_j} \frac {\partial \phi_k^\prime}{\partial \phi_j}-\frac {\partial \phi_i^\prime}{\partial \phi_j} \frac {\partial \phi_k^\prime}{\partial p_j})=0.$$ 
These $(2n^2-n)$ equations can be written in the more familiar matrix form,
\begin{equation}
MJM^T=J,
\end{equation}
where the $(2n \times 2n)$ matrices $M$ and $J$ are given by,
$$M=\left( \matrix{\partial p_i^\prime/\partial p_j &\partial  p_i^\prime/\partial \phi_j \cr
\partial \phi_i^\prime/\partial p_j &\partial \phi_i^\prime/\partial \phi_j 
\cr} \right), \ \ \ \ \ J=\left( \matrix{0  &\delta_{ij}\cr   
-\delta_{ij}  &0\cr} \right).$$
Equation (16) is recognized as expressing the necessary and sufficient condition (the symplectic condition \cite{Goldstein}) for the canonicality of the coordinate transformation (14); i.e.; the necessary and sufficient condition for the covariance of the Hamilton-like equations,
\begin{equation}
\dot{p_i}= \frac {\partial {\cal H}}{\partial \phi_i}, \ \ \ \ \dot{\phi_i}= -   \frac {\partial {\cal H}}{\partial p_i},
\end{equation}
under the transformation; ${\cal H}$ being a scalar
(${\cal H} (p_i,\phi_i,t)= {\cal H^\prime} (p_i^\prime,\phi_i^\prime,t)$)
with the dimensions of $time^{-1}$, of course. Adopting units $\hbar=1$, the Hamiltonian ${\cal H}$, clearly, should be identified with the mean (expectation) energy of the preparation which has the same value in all frames (representations). The canonical equations (17), being the only covariant set of equations under (14), then provide a unique candidate for the `equations of motion' of the preparation. Making contact with the standard formulation of quantum mechanics, we have,
$$\psi_i= \sqrt{p_i}\ e^{i \phi_i}, \ \ \ {\cal H} (p_i,\phi_i)=<\psi|H|\psi>= \sum_{ij} H_{ij} \sqrt{p_i p_j}\ e^{-i \phi_{ij}} ,$$
where $H$ is the Hamiltonian operator. Whence, (17) translates into,
$$i \dot{\psi_i}= \sum_{j} H_{ij} \psi_j,$$
which is just the Shr\"{o}dinger equation in the representation provided by the measurement apparatus; the covariance of (17) under the transformation (14) corresponds to the covariance of the latter under unitary transformations. The Shr\"{o}dinger equation, therefore, emerges naturally from the canonical property of (the transformation law of) the preparation space.

Now, it is always possible to work in the reference frame of the energy measurement apparatus, where the mean energy is given, in terms of the measurement results $\{E_i\}$, simply by ${\cal H}= \sum_{i} p_i E_i$. The coordinates $\phi_i$ are, therefore, `cyclic' and the equations of motion then yield,
$$p_i=const., \ \ \ \ \phi_i=-E_i t.$$
In the standard formulation, this solution corresponds to the superposition of energy eigenstates (stationary states), with components $\psi_i= \sqrt{p_i}\ e^{-i E_i t}$,  as the solution of the Shr\"{o}dinger equation in the energy representation. 

Due to the canonical property of its symmetry group, the preparation space, therefore, provides a new framework in which quantum mechanics appears in a manifestly canonical formulation. Notice how in this framework, the unitary transformation group of quantum mechanics on the one hand, and the Shr\"{o}dinger equation on the other, both emerge through the invariance property of the space. 

The dynamics in the preparation space, whence, closely resembles classical dynamics in the phase space picture. In particular, the canonically conjugate coordinates $(p_i,\phi_i)$ determine the evolution trajectory of a preparation in the preparation space of an isolated system with mean energy ${\cal H}$. On such trajectories, equations of motion (17) imply,
$$\dot{{\cal H}}= \sum_{i} (\frac {\partial {\cal H}}{\partial p_i}\ \dot{p_i} +\frac {\partial {\cal H}}{\partial \phi_i}\ \dot{\phi_i})=0,$$
i.e., ${\cal H}$ is a constant of motion; the mean energy of an isolated preparation, expectedly, does not change with time (provided, as we have assumed, the Hamiltonian does not depend explicitly on time). Furthermore, due to the time development of the preparation, the mean value of an arbitrary observable $F$, namely the scalar,
$$f(p_i,\phi_i) \equiv <\psi|F|\psi>=\sum_{ij} F_{ij} \sqrt{p_i p_j}\ e^{-i \phi_{ij}},$$
thus becomes a dynamical variable in the preparation space. Its dynamics follows from the equations of motion (17) to be determined from,
\begin{equation}
\dot{f}=\sum_{i} (\frac {\partial f}{\partial p_i} \frac {\partial {\cal H}}{\partial \phi_i}-\frac {\partial f}{\partial \phi_i} \frac {\partial {\cal H}}{\partial p_i}) \equiv \{f,{\cal H}\},
\end{equation}
where $\{f,{\cal H}\}$ denotes the Poisson bracket of $f$ and ${\cal H}$. Needless to say, because Poisson brackets are invariant under canonical transformations, the dynamics is independent of the choice of the reference frame of the measurement apparatus. Equation (18), of course, corresponds to the equation,
$$\dot{f}= \frac{1}{i} <[F,H]>,$$
of the standard formulation, as can be demonstrated directly.

Taking further advantage of the canonical formulation, in analogy with classical mechanics in phase space, the time evolution of an isolated preparation can be represented by a succession of infinitesimal canonical transformations generated by ${\cal H}$. Then, because the volume element of the preparation space remains invariant under canonical transformations ($ \delta(\sum_{i} p^\prime_i-1) d^n p^\prime\ d^n \phi^\prime=\delta(\sum_{i} p_i-1)\ \|M\| d^n p\ d^n \phi=\delta(\sum_{i} p_i-1) d^n p\ d^n \phi $), it follows that the volume element is a constant of motion. This could be used to obtain a Liouvill-like equation for the density of points in the preparation space, which would then serve as a basis for a new (canonical) formulation of quantum statistical mechanics; covariant under all (canonical) coordinate transformations in the space.

\section{Example: The preparation space of a two-level system}
The line element (3) can, also, be written as
$$ds^2=\sum_{i} \frac{dp_i^2}{4p_i^2}+\sum_{i<j} p_i p_j\ d\phi_{ij}^2.$$
Of the $\frac {1}{2} n(n-1)$ relative degrees of disturbance $\phi_{ij}$, only $(n-1)$ are independent; all $\phi_{ij}$ can be expressed in terms of, say, the $(n-1)$ values $\phi_{in}=\phi_i-\phi_n$. Indeed, the arbitrariness of the absolute values of the degrees of disturbance can always be used to set one of the $\phi_i$'s, say $\phi_n$, equal to zero which amounts to choosing their values relative to $\phi_n$. Bearing in mind the condition $\sum_{i} p_i=1$ too, the preparation space is, therefore, $2(n-1)$ dimensional. The description of a preparation, hence, involves $2(n-1)$ {\it independent} coordinates. Instead of viewing the preparation manifold extrinsically, i.e., as a hypersurface in the $2n$ dimensional enveloping space (in terms of the coordinates $(p_i,\phi_i)$), it can be dealt with intrinsically by a set of $2(n-1)$ coordinates. This approach turns out to be particularly interesting for a two level system ($n=2$), where the prepapration space is two dimensional. In this case,
$$ds^2=\frac{dp_1^2}{4p_1^2}+\frac{dp_2^2}{4p_2^2}+p_1 p_2\ d\phi_{12}^2.$$
Introducing the independent coordinates $(\theta,\phi)$ by,
$$p_1=\cos^2 \frac{\theta}{2}, \ \ \ p_2=\sin^2 \frac{\theta}{2},\ \ \ \phi \equiv \phi_{12},$$
where $0 \leq \theta \leq \pi$, the line element reduces to,
$$ds^2=\frac{1}{4} (d\theta^2+\sin^2\theta\ d\phi^2),$$
i.e., the preparation space is a sphere with radius $\frac {1}{2}$. The unitary transformations of quantum mechanics, thus, correspond to general rotations of the coordinate system $(\theta,\phi)$ on the sphere, which constitute the invariance group of the line element. Using conditions (12), the transformation law (14) for $p_i$ reduces to,
\begin{equation}
\cos \theta^\prime=\cos \alpha\ \cos \theta+ \sin \alpha\ \sin \theta\ \cos (\phi-\beta),
\end{equation} 
where ($0 \leq \alpha \leq \pi$), 
$$\cos^2 \frac{\alpha}{2} \equiv w_{11}=w_{22}, \ \ \ \sin^2 \frac{\alpha}{2}\equiv w_{12}=w_{21},\ \ \ \beta \equiv \beta_{11}-\beta_{12}.$$
Considering the aforementioned rotations, the above transformation formula, which is just the cosine law of spherical trigonometry, appears obvious: Let a given point $P$ be described by the coordinate sets $(\theta,\phi)$ and $(\theta^\prime,\phi^\prime)$ with respect to the origins $O$ and $O^\prime$, respectively, of two reference frames on the sphere. Then, the relation between the two coordinate sets can be obtained by just considering the spherical triangle $OO^\prime P$. In particular, if the coordinates of $O^\prime$ with respect to $O$ are denoted by $(\alpha,\beta)$, equation (19) will just correspond to the cosine law for this triangle.

Introducing $p=\frac{1}{2} \cos \theta$, the equations of motion reduce to,
\begin{equation}
\dot{p}= \frac {\partial {\cal H}}{\partial \phi}, \ \ \ \ \dot{\phi}= -\frac {\partial {\cal H}}{\partial p},
\end{equation}
so that $(p,\phi)$ form a canonically conjugate pair along the evolution trajectories on the sphere. Having established the sphericity of the preparation space by means of coordinates $(\theta,\phi)$, we can, henceforth, employ the canonical coordinates $(p,\phi)$, whose transformation law; being canonical; preserves the form of the equations of motion (20). Now, it is always possible to rotate to a coordinate system in which $\phi$ is cyclic; such a reference frame being provided by the energy measurement apparatus, where,
$${\cal H}=E_1 p_1+E_2 p_2= \frac{1}{2} (E_1+E_2)+(E_1-E_2)p.$$
The equations of motion (20), then yield,
$$p=const., \ \ \ \ \phi= -(E_1-E_2)t.$$
This, of course, corresponds to the familiar superposition of energy eigenstates (with relative phase given by $\phi$) in the standard formulation. The evolution trajectory of an arbitrary isolated preparation on the sphere is, therefore, circular ($\theta=const.$); the sense and angular velocity of the motion of the preparation point on the circle ($\dot {\phi}$) being determined by the difference in the energy levels.


\begin{thebibliography}{widest-label}
\bibitem{note} Incidentally, in the subjective view of probability, the same situation exists for the degrees of plausibility $\{p_i\}$, where the freedom in assigning absolute values is removed by the normalization condition $\sum_{i} p_i=1$.
\bibitem{note2} Such would be the case if preparations were to be specified with respect to measurement contexts by their $\{p_i\}$ alone.
\end{thebibliography}
\end{document}